\documentstyle[12pt,draft,epsf]{article}

\newcommand \be  {\begin{equation}}
\newcommand \ee  {\end{equation}}
\newcommand \bea {\begin{eqnarray} \nonumber }
\newcommand \eea {\end{eqnarray}}

\newcommand \al  {\alpha}
\newcommand \de  {\delta}
\newcommand \eps {\epsilon}
\newcommand \g   {\gamma}
\newcommand \la  {\lambda}
\newcommand \La  {\Lambda}
\newcommand \s   {\sigma}

\newcommand \cH  {{\cal H}}

\newcommand \NE {\not=}

\begin{document}

\title{On the Static and Dynamical Transition in the Mean-Field Potts glass}
\author{Emilio De Santis$^{(a)}$, Giorgio Parisi$^{(b)}$ and Felix Ritort$^{(a,b)}$\\[0.5em]
  {\small (a): Dipartimento di Fisica and Infn, Universit\`a di Roma
    {\em Tor Vergata}}\\
  {\small \ \  Viale della Ricerca Scientifica, 00133 Roma (Italy)}\\
  {\small (b): Dipartimento di Fisica and Infn, Universit\`a di Roma
    {\em La Sapienza}}\\
  {\small \ \  P. A. Moro 2, 00185 Roma (Italy)}\\[0.5em]}
\date{October, 1994}
\maketitle

\begin{abstract}
We study the static as well as the glassy or dynamical transition in
the mean-field $p$-state Potts glass. By numerical solution of the
saddle point equations we investigate the static and the dynamical
transition for all values of $p$ in the non-perturbative regime $p>4$.
The static and dynamical Edwards-Anderson parameter increase with $p$
logarithmically.  This makes the glassy transition temperature lie very
close to the static one. We compare the main predictions of the theory
with the numerical simulations.
\end{abstract}

\vfill

\baselineskip 6mm
\begin{flushright}
  {\bf  cond-mat/9410093}\\
  {\bf ROM2F/94/45}\\
\end{flushright}

\newpage

\section{Introduction\protect\label{S_INT}}

This work is devoted to the study of the glassy properties of the
mean-field Potts glass. Very recently there have been new developments
in the spin-glass theory concerning frustrated mean-field models without
explicit disorder \cite{I,II,BoMe,FrHe,CuKuPaRi,fully,MiRi}. It has been
shown that these systems do have a glassy transition temperature below
which thermal fluctuations are very small and dynamical relaxations are
very slow.  Even though these results are not new in the context of
disordered systems it is much interesting to know that non disordered
models also share these properties.

The purpose of this work is to study the glassy behavior of a
disordered spin glass. In general, these systems have a static
transition $T_{RSB}$ where replica symmetry is broken. The breaking of
the replica symmetry can occur in two ways. There can be a continuous
breaking pattern (as happens in the case of the Sherrington-Kirkpatrick
(SK) model \cite{SK}) or there can be a one step breaking of the replica
symmetry (as happens in $p$-spin models with $p>2$
\cite{GARDNER}). Also one can find intermediate phases where there is
a pattern with one step of breaking superimposed to a region with 
continuous breaking (as happens in $p$-spin models or Potts models at
low enough temperatures). The breaking pattern is fully described
by the order parameter $q(x)$ which is a function defined in the
interval $(0,1)$ \cite{Pa80}.

All these transitions are continuous from the thermodynamical point of
view, i.e. there is no latent heat. But in some cases they can be
first-order in the order parameter. This can occur because the
thermodynamic potentials are continuous (they are usually expressed as
integrals of the order parameter function $q(x)$).

The purpose of this work is the study of a disordered model with a
discontinuous transition in the order parameter. These systems
generally have a temperature $T_G$ where a dynamic instability
appears.  This temperature is called the glass temperature and is
higher than the transition $T_{RSB}$ where the replica symmetry
breaks. The first observation of this type was due to Kirkpatrick and
Thirumalai who solved the off-equilibrium dynamics for the $p$-spin
model above the glass temperature \cite{KiTh}.  Subsequently,
Kirkpatrick, Thirumalai and Wolynes studied the Potts mean-field glass
reaching similar conclusions \cite{KiWo,ThKi}. Similar results were
obtained in case of the $p$-spin spherical spin glass by Crisanti,
Horner and Sommers \cite{CrHoSo}. Below the glass transition it has
been shown by Cugliandolo and Kurchan \cite{CUKU} that the energy of
the $p$-spin spherical spin glass model in the low temperature phase
is higher than that predicted by the statics. 
For times larger than a
time scale (which diverges exponentially with the size of the system)
it is expected that the energy of the system will relax to its
equilibrium value. How fast this time scale grows with the size of the
system depends on particular features of the glass transition like the
discontinuity in the Edwards-Anderson parameter $q_G$.

In order to investigate the glassy behavior of a disordered model we
have decided to study the infinite-ranged Potts glass model. The
reason is threefold. On the one hand, in the Potts model $p$ is a
tunning parameter for the magnitude of the static and the dynamical
transition. On the other hand,
the Potts model is amenable of numerical tests while other models like
the $p$-spin model (Ising or spherical) are time consuming which makes
numerical simulations practically impossible for $p>3$. The situation
is different in case of the random orthogonal model \cite{II} where
the replica theory predicts the existence of a glassy phase in good
agreement with the numerical simulations.  Finally, the Potts glass
model lacks the reflection symmetry $\s_i\to -\s_i$ of some other
models. This makes it more similar to real structural glasses.

As we will see later the Potts glass model present one (not serious)
drawback. This is that for $p>2$ the systems order ferromagnetically
at low enough temperatures. In order to investigate the spin-glass
behavior it is necessary to introduce an additional antiferromagnetic
coupling constant. It is in these conditions that we have investigated
the glassy features of the discontinuous transition.  The research of
a glass transition in presence of ferromagnetic order remains an
interesting open problem.

This work completely solves the replica equations for the Potts model
for arbitrary number of states $p$. We will be able to exactly compute
the static and the dynamical transition and we will compare the
predictions with the numerical simulations.

We will see that a complete dynamical freezing never takes place, even
for very large values of $p$. For generic $p$ there is always a residual
entropy at the static transition $T_{RSB}$. As shown by Gross, Kanter
and Sompolinsky \cite{GrKaSo} in the limit $p\to\infty$ the statics of
the Potts model converges to the Random Energy Model (REM)
\cite{DERRIDA,GrMe}. We will see that the convergence of the statics of the
Potts glass model to the REM when $p\to\infty$ is very slow
(logarithmic in $p$).  Surprisingly, we will see that also the
dynamics converges logarithmically with $p$ to a fully frozen dynamics
but even more slowly than does the statics. For all practical
purposes, i.e. for reasonable values of $p$, the system is never fully
frozen. In addition we will see that, for $p>4$, the dynamical
transition (also called the glass transition) is always very close to
the static transition. This makes the glassy behavior of the Potts
model very different from other models with a discontinuous transition
in the order parameter like, for instance, the $p$-spin interaction
Ising spin-glass model where the static and the dynamic
Edwards-Anderson order parameter increase relatively fast with $p$.


This partial freezing which occurs for the mean-field Potts
glass has to be compared with deterministic
models without quenched disorder and disordered spin-glasses. On the
one hand, models like the low autocorrelation binary sequences
\cite{I}, fully frustrated lattices in the mean-field limit
\cite{fully} or discrete matrix models \cite{CuKuPaRi} display a stronger
glass transition because $q_G\sim 1$. In those cases there is no
quenched disorder and frustration alone is the responsible to the
existence of the glassy phase. The origin of the frustration is purely
dynamical and self-induced by the dynamical process \cite{BoMe}.  On
the other hand, disordered models like the $p$-spin glass model
\cite{GARDNER} or the random orthogonal model (ROM) \cite{II} do have
a stronger freezing at the dynamical transition temperature. The
presence of uncorrelated quenched disorder (i.e.
$\overline{J_{ij}J_{kl}}=\overline{J_{ij}}\,\,\overline{J_{kl}}$) in
the Potts glass, and also in the SK model, has the effect of softening
the discontinuous transition.  In some sense frustration corresponds
to the existence of some constraints on the different values of the
quenched couplings $J_{ij}$. This is what happens in the random
orthoghonal model where the $J_{ij}$ form an orthogonal random matrix.
Also in the case of the $p$-spin glass model the quenched disorder
variables $J_{i_1i_2...i_p}$ tend to frustrate the system much more
than in the Potts case. In the Potts glass case, for a fixed value of
the number of states $p$, the number of quenched variables goes like
$N^2$ ($N$ is the number of sites) while in the $p$-spin glass model
this number increases much faster with $N$ for $p>2$ (like
$N!/(N-p)!\sim N^p$ for finite $p$).

The work is divided as follows. In section 2 we introduce the model and
we write closed expressions for the free energy at first order of
replica symmetry breaking. In section 3 we solve
numerically the static equations at one step of replica symmetry
breaking and we determine the static and the dynamical transition.
Section 4 compares the predictions to the numerical simulations.
Finally we present the conclusions.

\section{Static replica equations for the Potts glass}

The Potts glass model is defined by the random Hamiltonian 

\be
\cH=-p\sum_{i<j}\,J_{ij}\delta_{\s_i\s_j}~~~~,
\label{eq1}
\ee

\noindent where $p$ is the number of states and the variables $\s$ can take the
values $0,1,..,p-1$.  The sum is extended over all $\frac{N(N-1)}{2}$
pairs in the lattice and $N$ is the number of sites. The couplings
$J_{ij}$ are randomly distributed with mean $\frac{J_0}{N}$ and variance
$\frac{1}{N}$. In order to solve this random model we apply the replica
method in order to compute the free energy $f$

\be
\beta f=lim_{n\to 0}\frac{\log \overline{Z_{J}^n}}{Nn}
\label{eq2}
\ee

\noindent where $n$ is the number of replicas and the overline means
average over the disorder. Performing the usual transformations
(avergaging over the disorder, decoupling the sites and introducing
auxiliary fields) and using the identity ($a,b=1,..,n$ are replica
indices)

\be
\delta_{\s_i^a\s_j^b}=\sum_{r=0}^{p-1}\,\delta_{\s_i^a\,r}
\delta_{\s_j^b\,r}~~~~~~,
\label{eq3}
\ee

\noindent one gets the following result 
\footnote {Alternatively one can use the simplex representation \cite{Zia}}

\be
\overline{Z_{J}^n}=\int\,dm_a^r\,dQ_{ab}^{rs}\,e^{-N\,A[m,Q]}~~~~,
\label{eq4}
\ee

\noindent where $r,s=0,...,p-1$ denote the Potts states. The function
$A[m,Q]$ is given by

\be
A[m,Q]=\frac{n\beta^2(1-p)}{4}+\frac{\beta^2}{2p}(J_0+\frac{p-2}{2})
\,\sum_{ar}(m_a^r)^2
+\frac{\beta^2}{2p^2}\sum_{a<b}\sum_{r,s}(Q_{ab}^{rs})^2-\log
Tr_{\s}\,e^{H[m,Q]}
\label{eq5}
\ee

with
\be
H[m,Q]=\frac{\beta}{p}(J_0+\frac{\beta(p-2)}{2})\sum_{a,r}\,m_a^r\,
(p\delta_{\s_a\,r}-1)\,+\,
\frac{\beta^2}{p^2}\sum_{a<b}\sum_{rs}\,Q_{ab}^{rs}\,(p\delta_{\s_a\,r}-1)
(p\delta_{\s_b\,s}-1)~~~~~.
\label{eq6}
\ee

\noindent The stationary saddle point equations read

\bea
m_a^r=\langle p\delta_{\s_a\,r}-1\rangle\\
Q_{ab}^{rs}=\langle (p\delta_{\s_a\,r}-1)(p\delta_{\s_b\,s}-1)\rangle
\label{eq7}
\eea

\noindent where the mean $\langle...\rangle$ is evaluated over the effective
Hamiltonian in eq.(\ref{eq6}).
The order parameters $m$ and $Q$ satisfy the constraints 
\bea
\sum_r\,m_a^r=0\\
\sum_r\,Q_{ab}^{rs}=0~~~~.
\label{eq8}
\eea

In the particular case $p=2$ with $Q_{ab}^{r\ne
s}=-Q_{ab},\,\,Q_{ab}^{rr}=Q_{ab}$ one recovers the solution for the
SK model \cite{SK}. It can be shown that ferromagnetic order is always
preferred for $p>2$ for low enough temperatures. An estimate $T_E$
for the temperature $T_F$ below which ferromagnetic order appears is given
by the following condition \cite{EdSh}
\be
J_0+\frac{p-2}{2T_E}=1~~.
\label{eqJ0}
\ee

In the special case $J_0=0$ the ferromagnetic transition appears below
$T=1$ for $p<4$ and above that temperature for $p>4$. Our main
interest in this work is the study of the spin-glass transition. In
order not to observe the ferromagnetic transition it will be necessary
to add an antiferromagnetic coupling in case $p>4$. This will be
discussed further at the end of this section. The spin-glass solution
is given by $m_a^r=0$. This means that all different $p$ states are
equally populated. The saddle point equations are
independent of $J_0$ and the replica symmetric solution in this case
is given by

\bea
Q_{ab}^{r\ne s}=-q\\
Q_{ab}^{rr}=q(p-1)~~~.
\label{eq9}
\eea

\noindent Substituting this result in eq.(\ref{eq7}) we obtain
\be
\beta f=\frac{\beta^2}{4}(1-p)(1-q)^2-
\int_{-\infty}^{\infty}\,\prod_{r=1}^p\,(\frac{dy_r}{\sqrt{2\pi}}e^
{-\frac{y_r^2}{2}})\,\log(\sum_{r=1}^p\,exp(\beta(qp)^{\frac{1}{2}}y_r))~~~.
\label{eq10}
\ee

The high-temperature result $q=0$ gives the free energy $f$, the
internal energy $u$ and the entropy $s$,

\bea
\beta f=\frac{\beta^2(1-p)}{4}-\log (p)\\
u=\frac{\beta(p-1)}{2}\\
s=\frac{\beta^2(1-p)}{4}\,+\,\log(p) .
\label{eq11}
\eea

Because the entropy has to be positive one finds that the replica symmetric
solution breaks down, at least above or equalt to

\be
T_0=\biggl (\frac{(p-1)}{4\log (p)}\biggr )^{\frac{1}{2}}~~~~.
\label{eq12}
\ee

It has been shown \cite{ErLa} that there is a continuous phase transition
at $T_c=1$ for $p<6$ which is unstable for $p\ge 2$. This transition
ceases to exist above $p=6$ and cannot be found within the replica
symmetric hypothesis.

It is necessary to break the replica symmetry.  By expanding the free
energy eq.(\ref{eq5}) close to $T_c=1$, Gross, Kanter and Sompolinsky
\cite{GrKaSo} have found two different regimes according to the value of 
$p$. In both cases the correct solution is given by one step of
breaking. In the region $2.8<p<4$ the transition is continuous. The
breaking parameter $m$ is $\frac{p-2}{2}$ at the transition
temperature $T_c=1$. At low enough temperatures the entropy of
the one-step solution becomes negative and a continuous breaking is then
necessary. In the regime $p>4$ the transition is discontinuous in $Q$
and the breakpoint parameter $m$ is equal to $1$ at the transition
temperature $T_c>1$. Cwilich and Kirkpatrick \cite{KiCw} have shown
that this one step solution is always stable for $p>p^{*}=2.82$ below
but close to $T_c$.

At first order of replica symmetry breaking we subdivide the $n$
replicas into $\frac{n}{m}$ blocks. Each block contains $m$ replicas
\cite{LIBRO}. The order parameter $Q_{ab}^{rs}$ takes a certain value
when both replicas $a,b$ belong to the same subblock and it is zero when
both indices belong to two different subblocks. More explicitely, if $K$
denotes a subblock, we impose

\bea
Q_{ab}^{r\ne s}=-q~~~~~~(a,b \in K),~~~~~Q_{ab}^{r\ne s}=0~~~~~~(otherwise)\\
Q_{ab}^{rr}=-q(p-1)~~~~~~(a,b \in K),~~~~~Q_{ab}^{rr}=0~~~~~~(otherwise)
\label{eq13}
\eea

We obtain the result,

\bea
\beta f=\frac{\beta^2}{4}(1-p)\,+\,\frac{\beta^2}{4}(m-1)(p-1)q^2\,+\,
\frac{\beta^2}{2} q(p-1)\,+\,\frac{\beta^2}{2}qm-\\
\frac{1}{m}\log
\int_{-\infty}^{\infty}\,\prod_{r=1}^p\,(\frac{dy_r}{\sqrt{2\pi}}e^
{-\frac{y_r^2}{2}})\,(\sum_{r=1}^p\,exp(\beta(qp)^{\frac{1}{2}}y_r))^m
\label{eq14}
\eea

The corresponding saddle point equations are
\be
\frac{\partial f}{\partial q}=\frac{\partial f}{\partial m}=0~~~~,
\label{saddle}
\ee
which determine the correct solution. It is possible to solve
pertubatively these equations in three different cases:

\begin{itemize}
\item Expanding around $p=4$ since in this limit case the transition is
quasi-continuous \cite{KiWo,ThKi}.  This technique has be applied also
in case of the $p$-spin model \cite{KiTh}.

\item Expanding around $T=1$ using the so-called effective approximation for
the free energy eq.(\ref{eq5}) up to order $Q^4$ \cite{KiCw}.

\item Solving the limit $p\to\infty$. In this limiting the model
converges to the random energy model \cite{Gold}.  For recent work see
\cite{Campellone}

\end{itemize}

We are interested in the glassy behavior of the Potts model. Our
approach will be to numerically solve the equation (\ref{saddle}).
This is the purpose of the next section.

Some comments are in order regarding the existence of the
ferromagnetic transition. We said previously that the system orders
ferromagnetically at low enough temperatures. The temperature $T_F$
below which the system orders ferromagnetically is smaller than $T_E$
with $T_E<1$ for $p<4$.  Also for $p<4$ the spin-glass transition
appears at $T_{RSB}=1$.  This means that in the regime $p<4$ the
spin-glass transition occurs at a temperature $T_{RSB}$ larger than
the temperature $T_F$ at which ferromagnetic order sets in. On the
other hand, for $p>4$ the spin-glass transition $T_{RSB}$ occurs at a
temperature greater than 1 but smaller than $T_F$. In order that
$T_F<T_{RSB}$ it is necessary to introduce a negative value for $J_0$.
In our numerical simulations we have chosen $J_0=\frac{4-p}{2}$ in
case $p>4$ and $J_0=0$ for $p<4$. In this way the spin-glass
transition occurs at a larger temperature than the ferromagnetic
ordering. Now, let us suppose that we perform a dynamical process of
the system in which the temperature is slowly decreased starting from
the high-temperature phase. We think that, once the system has entered
the metastable glassy phase, then it remains trapped in this phase for
a time which diverges exponentially with the size of the system.
Consequently, the system is unable to see the ferromagnetic transition
which occurs at a lower temperature. Two reasons reinforce this
observation:

\begin{itemize}
\item Numerical studies of the case $J_0=0$ for $p>4$ (see section 4) 
show that the ferromagnetic transition occurs at a temperature $T_F$
smaller than $T_E$ (eq.(\ref{eqJ0})). Then, for a generic
negative value of $J_0$ we can expect the ferromagnetic transition to
appear at a temperature much lower than $T_{RSB}$.

\item We can expect there exists a glass transition associated to the
static spin-glass one which probably occurs at a temperature $T_G$
larger than $T_{RSB}$.
\end{itemize}

In summary, by choosing $J_0$ as indicated above, the ferromagnetic
transition temperature will always be smaller than the freezing
temperature at which the spin-glass ordering appears. Only the case
$p=4$ could be a little tricky because the estimate for the
ferromagnetic transition $T_E$ and the spin-glass transition coincide,
but even in this case we have not observed in the numerical simulations
a strong magnetic ordering.

We should note that there are very few works devoted to the
study of the ferromagnetic behavior in the mean-field Potts glass and we
think it would be interesting to investigate it.

\section{The static and the dynamical transition}

In this section we are going to solve numerically eq.(\ref{eq14}). As is
usual in spin-glass theory we have to maximize the free energy as a
function of $q$ and $m$. We face the problem of computing the
$p$-dimensional integral

\be
I=\int_{-\infty}^{\infty}\,\prod_{r=1}^p\,(\frac{dy_r}{\sqrt{2\pi}}e^
{-\frac{y_r^2}{2}})\,(\sum_{r=1}^p\,exp(\beta(qp)^{\frac{1}{2}}y_r))^m
\label{eq15}
\ee

Because the solution of the replica equations involve a maximization in
the $(q,m)$ plane it is mandatory to compute $I$ with relatively high
precision. Because $I$ is a $p$-dimensional integral it can be computed
with the usual techniques only for $p$ not too large. The most easy
thing one can do is to divide the $p$-dimensional space into small
squares and use the Simpson algorithm or a similar one.  The computation
time grows as a power of $p$. This makes the calculation practically
impossible for $p>3$. However, if one exploits the fact that the
integration argument is invariant under the permutation of the $p$
indices $r=0,..,p-1$ then the integration region can be reduced to the
hyperplane $y_1<y_2<...<y_p$ where the $y_r$ denote the coordinates of
one point. In this way one can gain a factor $p!$ in the computation
time and we have been able to solve with good enough precision
up to $p=7$. These preliminary methods should be considered as checks for
the main computation.

We have been able to reduce the $p$-dimensional integral to a two
dimensional integral. In this way the problem is completely resoluble,
at least numerically. We use the identity,

\be
A^{m-1}=\frac{1}{\Gamma(1-m)}\,\int_0^{\infty}dx\,x^{-m}\,e^{-Ax}
\label{eq16}
\ee

\noindent where $\Gamma(x)$ is the Gamma function that is well defined
for $x>0$, i.e. $m<1$ as is the case once the analytic continuation
$n\to 0$ ($n$ is the number of replicas) has been done.

We decompose the integrand in eq.(\ref{eq15}) as a product of two terms
$A*A^{m-1}$ with $A$ given by,

\be
A=\sum_{r=1}^p\,exp(\beta(qp)^{\frac{1}{2}}y_r)
\label{eq17}
\ee

Applying eq.(\ref{eq16}) and using the fact that the integrand in
eq.(\ref{eq15}) is invariant under permutation of the indices we get the
final result,

\be
I=\frac{pe^{(\frac{\beta^2qp}{2})}}
{\Gamma(1-m)}\,\int_0^{\infty}dx\,x^{-m}\,(w(x))^{p-1}
w(xe^{\frac{\beta^2qp}{2}})
\label{eq18}
\ee

where $w(x)$ is given by

\be
w(x)=\int_{-\infty}^{\infty}\frac{dy}{\sqrt{2\pi}} 
exp(-\frac{y^2}{2}-x\,\exp(\beta
(qp)^{\frac{1}{2}}y))\\
\label{eq19}
\ee

The integral over $x$ is well defined and free of divergences. However
one has to be careful evaluating the integrand close to $x=0$. We have
been able to maximize the free energy and completely solve the static
replica equations up to $p=40$.

The results are shown in figures 1 and 2 where we plot the variational
parameters $q$ and $m$ as a function of $T$. We plot the solutions for
the cases, $p=3,5,7,10,20,40$. The transition temperature also grows
with $p$. The solution of the integral eq.(\ref{eq18}) presents some
problems of precision at very low temperatures and also close to the
transition temperature where it is difficult to precisely determine
the value of the discontinuity.  A more precise way to compute the
critical temperature and the discontinuous jump of $q$ will be
presented below.  It is interesting to note how much slow is the
convergence to the random energy model as $p$ increases.  When $p$
increases the value of $q$ at the transition point grows very slowly.
In fact, in the limit $p\to\infty$, the value of $q$ converges to 1
and the entropy is zero at the transition temperature.  Using
eq.(\ref{eq12}) we obtain that the critical temperature grows like
$T_0$ of eq.(\ref{eq12}). This result was already noted in
\cite{GrKaSo}.

We have already observed that at very low temperatures the entropy of
the one-step solution becomes negative (of order $10^{-2}$).
Continuous breaking is necessary (as noted in \cite{GrKaSo}) but we
have not studied this type of solution. It is not clear to us if any
effect of this new transition could be observable in a numerical
simulation.  

We want to show now a more precise computation of the critical
temperature $T_{RSB}$ and the glass transition $T_G$. From the
dynamical point of view an instability in the dynamical equations
appears at a temperature $T_G$ above the static transition
$T_{RSB}$. Using the static approach, this dynamical temperature $T_G$
can be determined computing the smallest eigenvalue in the stability
matrix. The vanishing at $T_G$ of this eigenvalue, sometimes called the
replicon, corresponds to the marginality condition \cite{MARG}.  In
principle, this condition correctly determines the dynamical or glass
transition.  Anyway it is not clear if it is the correct description
of the dynamical behavior in the low temperature phase. This condition
has been numerically solved in the random orthoghonal model and it has
been shown that it correctly describes the dynamical energy below the
glass transition and not too low temperatures \cite{II}. It also
describes correctly the glass transition in case of deterministic
models. The interested reader is referred to \cite{II} for more
details. In order to determine the glass transition for the Potts case
we should compute the stability matrix of the problem. This is an
involved task (which has beeen done by Cwilich and Kirkpatrick close
to $T_c$\cite{KiCw}) and we will follow a different strategy (already
noted by Cwilich and Kirkpatrick but not fully explained). It can be
shown that in the limit $m\to 1$ the replicon eigenvalue coincides
with the longitudinal eigenvalue.  This result can be shown using the
exact expressions for the spectrum of the stability matrix which have
been reported in the literature at first order of replica symmetry
breaking \cite{Brunetti}.  From the stability analysis results of
Cwilich and Kirkpatrick this can also be directly tested in the Potts
glass case. Consequently, in order to determine the dynamical
transition, suffices to impose the marginality condition for the
longitudinal fluctuations.

We expand the free energy eq.(\ref{eq14}) around $m=1$,

\be
\beta f=\frac{1}{4}\beta^2(1-p)-\log(p)+(m-1)\biggl (
\frac{1}{4}\beta^2(p-1)q^2\,+\,\frac{1}{2}\beta^2 q(p+1)\,+\,\log(p)-
I_2\biggr )
\label{eq20}
\ee

\noindent where the integral $I_2$ is given by,

\be
I_2\,=\exp(-\frac{\beta^2pq}{2})\,
\int_{-\infty}^{\infty}\,\prod_{r=1}^p\,(\frac{dy_r}
{\sqrt{2\pi}}e^{-\frac{y_r^2}{2}})\,e^{\beta \sqrt{qp}y_1}
\log(\sum_{r=1}^p\,exp(\beta(qp)^{\frac{1}{2}}y_r))
\label{eq21}
\ee

For $m=1$ eq.(\ref{eq20}) reduces to the high-temperature free energy which
is independent of $q$. More generally, we can write the free energy as

\be
f=f_0+(m-1)f_1+O((m-1)^2)
\label{eq22}
\ee

where $f_0$ is independent of $q$. This general expansion locates the
static and the dynamic transition. For the static transition we look
for a temperature at which there is a solution $q_{RSB}$ which satisfies

\bea
\biggl (\frac{\partial f}{\partial q}\biggr )_{q=q_{RSB}}=\biggl (
\frac{\partial f_1}{\partial q}\biggr )_{q=q_{RSB}}=0\\
(f_1)_{q=q_{RSB}}=0
\label{eq23}
\eea

For the dynamical transition the stability is marginal and the second
derivative of $f$ respect to $q$ vanishes,

\bea
\biggl (\frac{\partial f}{\partial q}\biggr )_{q=q_G}=\biggl (
\frac{\partial f_1}{\partial q}\biggr )_{q=q_G}=0\\
\biggr (\frac{\partial^2 f}{\partial q^2}\biggr )_{q=q_G}=
\biggl (\frac{\partial^2 f_1}{\partial q^2}\biggr )_{q=q_G}=0~~~~.
\label{eq24}
\eea

The last equations correspond to the case in which an extremal solution
of eq.(\ref{eq20}) with $q_G\ne 0$ dissappears. It is then clear that
the dynamical transition temperature is always higher than the static
one.  We have solved these equations for different values of $p$. Now we
face the problem of computing the $p$-dimensional integral $I_2$. It can
be reduced to a two dimensional integral using the representation,

\be
\log(1+A)\,=\,\int_0^{\infty}\,\frac{dx}{x}e^{-x}(1-e^{-Ax})~~~~~.
\label{eq25}
\ee

\noindent and taking

\be
A\,=\,(\sum_{r=1}^p\,exp(\beta(qp)^{\frac{1}{2}}y_r))-1
\label{eq26}
\ee

\noindent we obtain the result

\be
I_2\,=\,\int_0^{\infty}\,\frac{dx}{x}e^{-x}\,\lbrace
1-e^{x}w(xe^{\frac{\beta^2qp}{2}})
w^{p-1}(x)\rbrace~~~~~~~~~.
\label{eq27}
\ee

with the same function $w$ as given in eq.(\ref{eq19}).  We have
solved equations eq.(\ref{eq23}) and eq.(\ref{eq24}) for different
values of $p$. Our results are summarized in Table 1. We find for each
value of $p$ two temperatures. One is $T_{RSB}$ and corresponds to the
static transition with the discontinuous jump $q_{RSB}$. The other one is
$T_G$ and corresponds to the dynamical transition with the
discontinuous jump $q_G$. Our results for the static transition are in
agreement with those found with the previous analysis using the
maximization procedure for the free energy. This is a check of our
procedures. Moreover this analysis provides a much more
precise determination of the values of $q_{RSB}$, $q_G$ and the
transition temperatures.

The results we have found are also in agreement with those reported by
Cwilich and Kirkpatrick, the only difference is that all their
computations are perturbative whereas ours are exact. As was obtained
in \cite{KiWo} and \cite{KiCw} one finds that $q_G/q_{RSB}=3/4$ for
$p$ close to 4. Looking at table 1 the reader can observe that the
ratio $q_{G}/q_{RSB}$ stays so close to $3/4$, even for large values of
$p$, that one is tempted to conclude that this is exact at all orders
in perturbation theory. Our numerical precision to solve the equations
(\ref{eq23}),(\ref{eq24}) is good enough to exclude this possibility.
From the results shown in the Table 1 it is clear that the convergence
to the $p\to\infty$ limit is very slow.  Fortunately, our numerical
program which solves the equations (\ref{eq23}) and (\ref{eq24}) is
enough accurate to explicitly show this slow convergence even for
exponentially large values of $p$.  We have solved the full equations
up to $p=10^6$. The results for $q_{RSB}$ and $q_G$ as a function of
$\frac{1}{\log(p)}$ are shown in figure 3. 
 
Furthermore in the Potts case the ratio $T_G/T_{RSB}$ grows very
slowly with $p$ being always smaller than $1.13$ up to $p=10^6$. The
proximity of the temperatures makes it difficult to discern one from
the other in numerical simulations. This proximity of the static and
the dynamic transition temperatures is very probably related to the
small value of the dynamical order parameter $q_G$ for large values of
$p$. From these results we expect the glassy behavior of the Potts
glass to be very different from other disordered spin-glass models.

For instance, in case of the $p$-spin interaction spin-glass model we
have also solved the equations corresponding to eq.(\ref{eq23}) and
eq.(\ref{eq24}).  We have found that both the static order parameter
$q_{RSB}$ and the dynamical $q_G$ converge to 1 in the limit
$p\to\infty$ much faster than the Potts case in agreement with
theoretical expansions around the $p\to\infty$ limit \cite{Campellone}.
For $p=3$ (the smallest value of $p$ compatible with a discontinuous
transition) one finds in the $p$-spin model,

\bea
q_{RSB}\simeq 0.81~~~~~ (T_{RSB}\simeq 0.65)\\
q_G\simeq 0.68~~~~~ (T_G\simeq 0.68)~~~.
\label{resp3}
\eea

For this particular model, the ratio $q_{G}/q_{RSB}$ tends to 1 in the limit
$p\to\infty$ and the ratio $T_G/T_{RSB}$ increases with $p$ much faster
than the Potts model does (for $p=10$ we find $q_G/q_{RSB}\simeq .97$ and
$T_G/T_{RSB}\simeq 1.38$)

In the next section we shall compare all these predictions with Monte
Carlo numerical simulations. We will see that the Potts glass transition
is always present but it is far from being a complete thermodynamic
freezing as happens in models where frustration is stronger (see, for
instance the random orthogonal model \cite{II}).  Before showing our
Monte Carlo results for the spin-glass transition it will be interesting
to present some results on the ferromagnetic ordering that takes place
in the Potts glass. The problem of the existence also of a real glass
transition above the ferromagnetic transition still remains open. Our
main interest is to show that if one does not introduce an
antiferromagnetic coupling then the ferromagnetic ordering takes place
even though the transition temperature is well below that given by
eq.(\ref{eqJ0}).

\section{Monte Carlo tests of the glass transition}

In order to simulate the Potts glass we have considered the Hamiltonian,

\be
\cH=-\sum_{i<j}\,J_{ij}(p\delta_{\s_i\s_j}-1)
\label{eq28}
\ee

The $J_{ij}$ are distributed with mean $J_0/N$ and variance equal to
$\frac{1}{N}$.  For computational reasons we have chosen a binary
distribution where the $J_{ij}$ can only take the values $\pm
\frac{1}{\sqrt{N}}$. The only difference between the Hamiltonians of
eq.(\ref{eq28}) and eq.(\ref{eq1}) is a constant which vanishes in the
thermodynamic limit. We have chosen this second version because we
have found that the addition of the constant strongly reduces the
sample to sample fluctuations in the high-$T$ region.  This should not
make too much difference for small values of $p$ but is crucial for
large values of $p$.  All simulations implement the Metropolis
algorithm with random updating.

The results we present in the next subsections correspond to
annealings in which we compute the main thermodynamic observables.
Starting from the high-temperature region the temperature is
progressively decreased. Statistics is collected at each temperature
and the time we stay at each temperature is the same for all
temperatures during the cooling procedure. We have computed the
internal energy, the magnetization of the different $p$-states and the
associated dissipative quantities, i.e.  the specific heat and the $p$
different magnetic susceptibilities corresponding to each one of the
$p$-states. The specific heat and the magnetic susceptibility of one
of the $p$ states is computed measuring the fluctuations of the
internal energy and the magnetization (see eq.(\ref{eq7}))
respectively.  Typically we performed several thousands of Monte Carlo
sweeps at each temperature. We have to call the attention of the
reader that our results are dependent on the time schedule of the
annealing only for very large values of $p$ (i.e. where the
finite-size corrections are large). Otherwise, one cannot observe a
sensible dependence of the different quantities on the time the system
stays at each temperature during the cooling procedure.  At least,
this dependence is of the same order as that arising from the sample
to sample fluctuation. Obviously we have performed annealing schedules
as large as possible within our computing capabilities. We will
eventually show the dependence on the annealing time in the large $p$
case.

\subsection{Ferromagnetic ordering with $J_0=0$}

When $J_0=0$ the system orders ferromagnetically. We have investigated
the ferromagnetic ordering for $p=10$. This value of $p$ is in the
regime ($p>4$) where the ferromagnetic transition is expected to
appear at a temperature higher than the spin-glass transition. From
eq.(\ref{eqJ0}) we expect ferromagnetic order to be present below
$T_E=4$. Figure 4 shows the internal energy as a function of the
temperature compared to the energy of the spin-glass phase and the
high-temperature result eq.(\ref{eq11}).  The energy is lower than
that corresponding to the spin-glass solution.  Figure 5 shows the
result for the magnetic susceptibility averaged over the different
$p=10$ states.  The first well defined peak is at $T_F=2.3$ which
corresponds to the temperature at which the energy of the system
departs from the high-temperature result (dashed line in figure 4).
The specific heat also shows a peak at that temperature.  The reader
will immediately recognize that temperature as the ferromagnetic
ordering temperature $T_F$ which is much lower than the estimate
$T_E$. It is interesting to note in figure 5 the emergence of further
peaks at lower temperatures.  These are a sign of new transitions in
the ferromagnetic phase. From the measurements of the magnetization we
have observed that at the transition temperature there is one state
which acquires a magnetization greater than zero which means that one
state is macroscopically populated. We also have observed that the
other peaks at lower temperatures correspond to the emergence of new
states which start to be macroscopically populated.

We have also studied the zero temperature ground state following a
steepest descent procedure. We have searched for stable configurations
against {\em one-spin} flip (for a generalization to stability against
{\em n-spin} flips see \cite{MOORE}). Starting from a random initial
configuration we sequentially move on the lattice selecting (among the
$p-1$ possibilities) the state of the variable $\s(i)$ which releases
the largest amount of energy.  In this way the system reaches a
metastable state that should be magnetized if the ground state is
ferromagnetic. We have repeated this procedure several times saving
the energy and the magnetization of the final configurations.  Figures
6 and 7 show the distribution probability of the energies and the
magnetization eq.(\ref{eq7}) of the stable configurations against {\em
one-spin} flip movings for the same model $J_0=0$, $p=10$ with
$N=100$.  Figure $6$ shows that the energies of this class of
metastable states are distributed very similarly to the form predicted
for the SK model \cite{BrayMoore}. This is a consequence of the glassy
nature of the ferromagnetic phase. We have verified that the minimum
energy found by the algorithm is higher than the energy obtained doing
a slow cooling starting from the high-temperature phase. This is a
proof of the glassy nature of this phase.  The ferromagnetic nature
(but glassy) of the phase is explicitely shown in figure figure 7. The
value of $m$ ranges from $m=-1$ to $m=p-1=9$. The peak at $m=-1$ is
consequence of the fact that only some of the $p=10$ states are
populated.

\subsection{The continuous transition ($p=3$)}

The case $p=3$ is indeed very similar to the SK model ($p=2$). Because
the transition is continuous the system relaxes very close to the true
energy. As mentioned in section 2, it now suffices to take $J_0=0$. In
this way the ferromagnetic transition lies well below the spin-glass
transition.  In fact, we have not observed any tendency of the system to
be magnetized at low temperatures. Figure 8 shows the internal energy as
a function of the temperature along with the one-step solution and the
high-temperature result eq.(\ref{eq11}).  Below the critical temperature
$T_{RSB}=1$ the data departs from the prediction.  Precisely at $T=1$
the specific heat and the magnetic susceptibility have a cusp.
Similar results for the internal energy were obtained for $p=4$.

\subsection{The discontinuous transition}

To investigate the discontinuous spin-glass transition we have chosen
$J_0=\frac{4-p}{2}$ for $p>4$. In this way the system first enters the
metastable glassy phase in which there is no ferromagnetic ordering.
In all our simulations we have observed that this is what happens and
that there is no tendency for the ferromagnetic domains to grow as the
temperature is decreased. At high temperatures the size of the domains
(i.e. the fraction of sites of the lattice which are in the same
state) is $1/p$. This is true down to very low temperatures where in
the worst case the size of the domains increase approximately ten
percent.  To make any tendency to the ferromagnetic ordering
completely dissapear we can increase the intensity of the
antiferromagnetic coupling. This is only possible if $p$ is not too
large because otherwise finite-size corrections (an consequently
finite-time effects) considerably increase.  In the regime of large
values of $p$ one can neglect finite-size effects only if $p\ll N$.
This is a problem of the simulations in the large $p$ regime (the
region where the glass transition can be clearly appreciated). We will
explicitly show the time dependence of the results of annealing for
$p=20$. We divide our results for the discontinuous transition in two
parts, the small and the large $p>4$ regime (corresponding to the
magnitude of the finite-size corrections).


\subsubsection{The small $p>4$ regime} 

We have measured the internal energy as a function of the temperature
for cases $p=5,10$. In this regime we have observed that the results
do not vary too much depending with the time schedule of the cooling
procedure. Comparison with theory is shown in figures 9 and 10.  For
these small values of $p$ the dynamical transition practically
coincides with the static one. Comparing to the previous continuous
case $p=3$ we see that the energy in the low $T$ region is very close
to the predicted one for $p=5$ and remains slightly above the expected
one in case $p=10$. This is the glassy phase where the system remains
trapped making excursions between several metastable states of similar
energy but without reaching the static phase of slightly lower free
energy. The difference in free energy (and energy) between the static
phase and the metastable glassy phase is small for $p=5$ and increases
with $p$.  It is important to note that the energy we are computing is
purely dynamical.
For $p\le 4$ this difference of free energy does not exist. This does
not mean that the system relaxes to the true ground state energy in an
annealing process (see figure $8$). In fact, for a continuous transition
we expect the system should relax to the static free energy at a finite
temperature very slowly (as a power law) very similarly to the
relaxation of the remanent energy or the remanent magnetisation in the
$SK$ model \cite{PaRi,Fe}. For a discontinuous transition the
relaxation of the free energy takes place also very slowly but to a
dynamical value higher than that predicted by the static approach. We
have also computed the specific heat and the magnetic susceptibility.
They display a cusp located approximately at the static transition
(and, because of its proximity, also the dynamical transition). The 
reader may be a little puzzled by the
data shown in figure 10 because
the dynamical energy departs form the high-$T$ behavior at a temperature
higher than the glass transition temperature. We think this occurs
because the static and the glass transition temperatures are very
close to each other. We will return to this point in the conclusion.

\subsubsection{The large $p>4$ regime} 

Finite-size corrections are important and one has to simulate large
sizes in order to reduce these effects.  We present the results of
annealings for $p=20,40$ in figures 11 and 12.  We decided to simulate
the Gaussian $J_{ij}$ model instead of the binary $\pm J$ one in order to
reduce finite-size corrections. As $p$ increases the finite-time effects
also increase and we have found a clear dependence of our results on the
cooling procedure. Figure 11 shows simulation results for $p=20$ for
$N=2000$ and two different cooling procedures. The simulation results
show a drift with the annealing time. For $p=40$ (figure 12) we show
simulations of two different sample realizations.  Since sample to
sample fluctuations increase with $p$, the relative magnitude of the
fluctuations of figure 12 should be considered as an upper bound for the
previous figures with smaller values of $p$. Also finite-time effects
are large for $p=40$ . From figures 11 and 12 it can be appreciated that
also in this case the energy departs from the high-$T$ result at a
temperature higher than predicted for the glass transition.

Glassy effects are much more pronounced in the large $p$-regime, the
dynamical energy being larger than the static one. All dissipative
quantities show a cusp very close to the dynamical transition. Even
though the static and the glass transition are very close one to the
other the fact that the energy of the sytem is much higher
than the static one (when approaching the glass transition) is a proof
that the system has entered the glassy phase.

\section{Conclusions}

We have studied the glassy behavior of the mean-field Potts glass. We
have been able to numerically solve the static equations at
first order of replica symmetry breaking. Also we have introduced a
simple method, already observed by Cwilich and Kirkpatrick \cite{KiCw},
which allows a full computation of the static and the dynamical or glass
transition and the associated Edwards-Anderson parameter.

We have numerically computed the parameters of the transition for
different values of $p$. We observe that the Edwards-Anderson
parameter at the glass transition $q_G$ increases logarithmically with
$p$.  The ratio of the static and glass temperature is smaller than
$1.13$ up to $p=10^6$. The situation is very different from other
disordered models such as the $p$-spin Ising model.  For that model
the dynamic transition temperature is much higher than the static one.

All our numerical results seem to indicate that the dynamical transition
takes place at a temperature higher than that predicted by the theory.
But this is due to the proximity of the dynamical transition to the static
one. If the dynamical transition temperature were much larger than the
static one then we would expect that the energy departs from the
high-$T$ result precisely at the dynamical temperature.  This is indeed
the situation one observes in low autocorrelation models
\cite{I,BoMe,MiRi} , in the random orthogonal model \cite{II,fully} and
discrete matrix models \cite{CuKuPaRi} . We would also expect this
situation for the $p$-spin Ising model in the regime
of not too small $p$, but unfortunately we are only able to perform
simulations for $p=3$.

For large values of $p$ (like $p=20,40$) we have observed a clear
dependence on the time spent during the cooling procedure . The origin
of the finite-time effects is related to the finite-size effects we
also observe in this regime. We expect that simulations for larger
sizes should give results nearly independent of the annealing time
leaving only a small thermalization time effect close to the glass
transition where critical effects begin to be important. We interpret
this effect in the following scenario.

There are two characteristic relaxation times in the system. The first
time $\tau_G$ diverges as the dynamical transition is approached, the
other one $\tau_s$ diverges as the static transition is approached.
Because $T_G$ is so close to $T_{RSB}$ the systems feels the static
low temperature phase very close to the dynamical transition
temperature. Above the glass temperature we have $\tau_G\sim\tau_s$
which is certainly large if the system is entering the low temperature
phase. Because the characteristic time scale $\tau_G$ increases very
fast only very close to the dynamical transition temperature then we
expect that close to $T_G$ the correlation time $\tau_s$ will set the
characteristic time scale above which our simulation results should be
time independent. Only for times larger than $\tau_s$ (which we are
not able to reach in our simulations) the system would behave as
dynamics predicts.  It is then clear that all our simulation results
are smeared by the static relaxation time $\tau_s$.  In the other
models mentioned in the previous paragraph the dynamical transition
temperature is much larger than the static one. Approaching the
dynamical transition the system is in the high temperature phase where
the relaxation time is very small.  Consequently, $\tau_G$ grows very
much only very close to $T_G$ (very probably diverges like $\tau_G\sim
(T-T_G)^{-\gamma}$ where $\gamma=2$ the typical value for mean-field
models \cite{MiRi}) and the system departs from the high-$T$ result
very close to that temperature.

Because the Potts model is only partially frozen at the glass transition
this is a model appropiate for study of the dynamics in the metastable
glassy phase. We expect that jumps among states could be seen
numerically without special effort. For models without disorder, the
system freezes quickly at the glass temperature and more involved
numerical techniques are needed in order to allow the system to change
state \cite{Mezard}. 


We would also like to draw attention to the interest of studying the
ferromagnetic ordering for zero mean coupling.  It would be
interesting to understand the static as well as the dynamical behavior
in that case.

\section*{Acknowledgements}

We are grateful to M. Campellone and E. Marinari for fruitful
discussions and D. Lancaster for a careful reading of the manuscript.
F. R has been supported by the INFN.

\vfill
\newpage
{\bf Figure caption}
\begin{itemize}

\item[Fig.~1] The one-step breaking parameter $q$ as a function of the
temperature. From left to right: $p=3,5,7,10,20,40$

\item[Fig.~2] The one-step breaking parameter $m$ as a function of the 
temperature. The different lines that intersect the upper horizontal
axis $m=1$ correspond from left to right to: $p=3,5,7,10,20,40$

\item[Fig.~3] The static and the dynamic Edwards-Anderson parameter as
a function of $\frac{1}{\log(p)}$. They increase logarithmically with
$p$. The dots are for the static value, the crosses for the dynamical
one.

\item[Fig.~4] Energy versus temperature for the case $p=10$ with
$J_0=0$. The continuous line corresponds to the one step spin-glass solution
and the dashed line is the high-$T$ result. The ferromagnetic transition
is close to $2.3$. Simulation results are for one sample and $N=1000$.
 
\item[Fig.~5] Magnetic susceptibility versus temperature for the case
$p=10$ with $J_0=0$. The first peak appears at $T_F\sim 2.3$.

\item[Fig.~6] Probability distribution of the energy of the ground
states for case $p=10$ and $J_0=0$ and $N=100$. 

\item[Fig.~7] Probability distribution of the magnetization associated to
the zero-temperature metastable states. The singularity at $m=-1$ means
that only some states are populated. The same parameters as in figure 6.

\item[Fig.~8] Energy versus temperature in case $p=3$, $J_0=0$. 
The continuous line is the one-step solution. The dashed line is the
high-$T$ result.  The transition temperature is $T_{RSB}=T_{G}=1$. The
full dots are for one sample and $N=2000$

\item[Fig.~9] Energy versus temperature in case $p=5$, $J_0=-1$. The continuous
line is the one-step solution. The dashed line is the high-$T$ result.
The transition temperatures are $T_{RSB}\simeq 1.0091\,,T_{G}\simeq
1.01$.  The full dots are simulation results for $N=1000$.

\item[Fig.~10] Energy versus temperature in case $p=10$, $J_0=-3$. 
The continuous line is the one-step solution. The dashed line is the
high-$T$ result.  The transition temperatures are $T_{RSB}\simeq
1.13\,,T_{G}\simeq 1.14$.  The full dots are simulation results for
$N=1000$.

\item[Fig.~11] Energy versus temperature in case $p=20$, $J_0=-8$.
The continuous line is the one-step solution. The dashed line is the
high-$T$ result.  The transition temperatures are $T_{RSB}\simeq
1.36\,,T_{G}\simeq 1.39$.  The crosses and the full dots correspond to
the Gaussian model with $N=2000$ and two cooling procedures (the
simulations with crosses are 10 times larger in simulation time than
the dots).

\item[Fig.~12] Energy versus temperature in case $p=40$, $J_0=-18$. 
The continuous line is the one-step solution. The dashed line is the
high-$T$ result.  The transition temperatures are $T_{RSB}\simeq
1.71\,,T_{G}\simeq 1.76$.  The crosses and the full squares correspond
to the Gaussian model with $N=1000$ and two different samples.

\end{itemize}
\vfill\eject

\begin{table}
\centering
\begin{tabular}{|c|c|c|c|c|c|c|}\hline
  & & & & & & \\
       $p$ &    $T_{RSB}$   &  $q_{RSB}$ &    $T_G$   &  $q_G$ 
& $T_G/T_{RSB}$ & $q_G/q_{RSB}$  \\ \hline
  & & & & & &\\
 $3$ & 1 & 0 & 1 & 0 &---&---  \\\hline 
  & & & & & &\\
 $4$ & 1 & 0 & 1 & 0 &---&---  \\\hline 
  & & & & & &\\
 $5$ & 1.0091 & 0.130 & 1.0100 & 0.0985 & 1.001 & 0.757 \\\hline 
  & & & & & &\\
 $7$ & 1.053 & 0.308 & 1.058 & 0.231 & 1.004 & 0.75 \\\hline 
  & & & & & & \\
 $10$ & 1.1312 & 0.452 & 1.142 & 0.328 & 1.009 & 0.725  \\\hline 
  & & & & & &\\
 $20$ & 1.364 & 0.641 & 1.393  & 0.468 & 1.02 & 0.73   \\\hline 
  & & & & & & \\
 $40$ & 1.711 & 0.752 & 1.765 & 0.551 & 1.03 & 0.732  \\\hline 
  & & & & & &\\
 $100$ & 2.388 & 0.838 & 2.496 & 0.633 & 1.045 & 0.755  \\\hline 
  & & & & & &\\
 $10^3$ & 6.075 & 0.931 & 6.51 & 0.721 & 1.07 & 0.774  \\\hline 
  & & & & & &\\
 $10^4$ & 16.54 & 0.966 & 18.05 & 0.769 & 1.091 & 0.796  \\\hline 
  & & & & & &\\
 $10^5$ & 46.69    & 0.981    & 51.64 & 0.802 & 1.1 & 0.817  \\\hline
  & & & & & &\\
$10^6$ & 134.65   & 0.989    & 150.5 & 0.835 & 1.12 & 0.844  \\\hline
\end{tabular}
\end{table}
\vskip0.7truecm
\centerline{Table 1.}
\vfill\eject

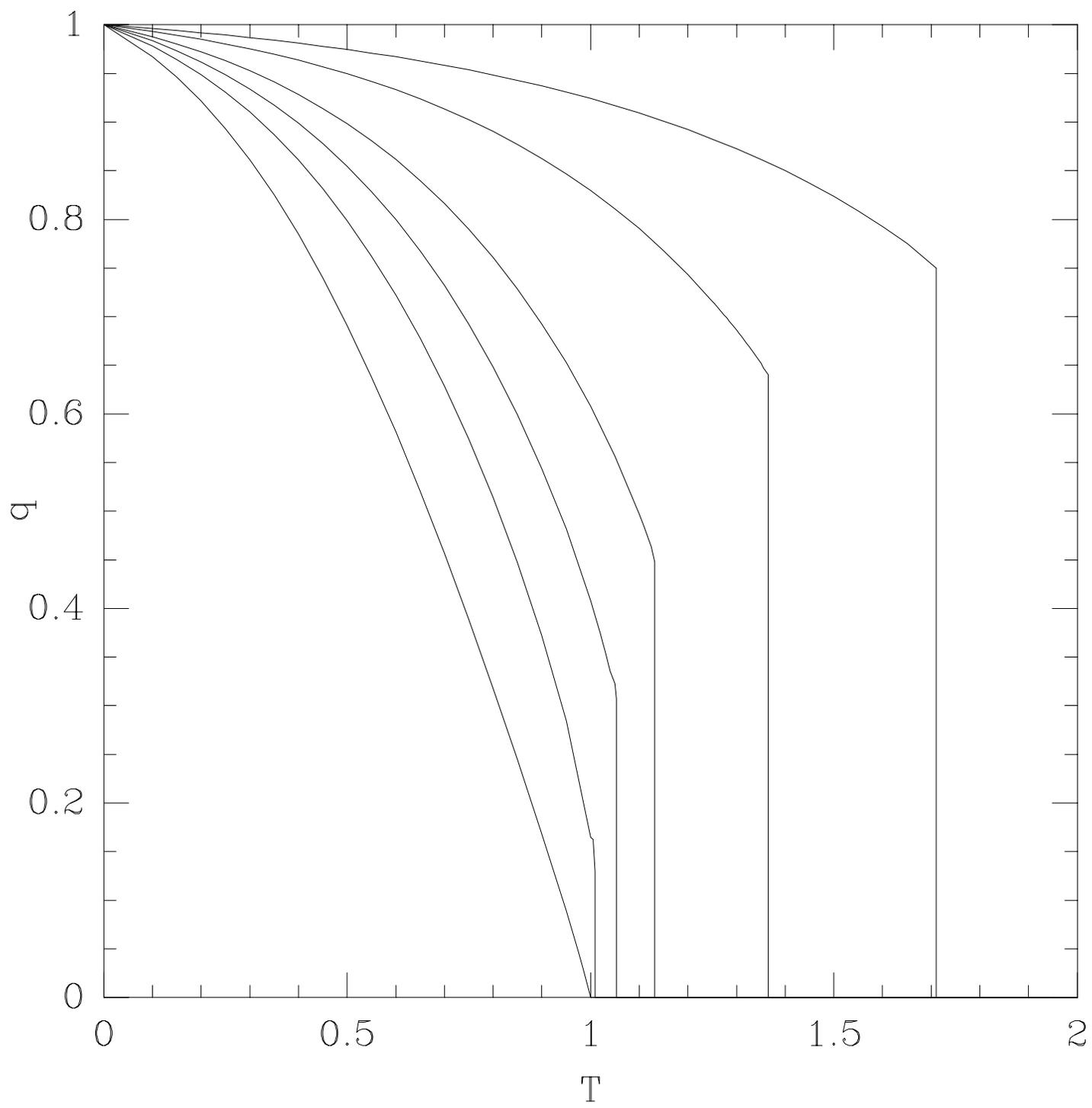
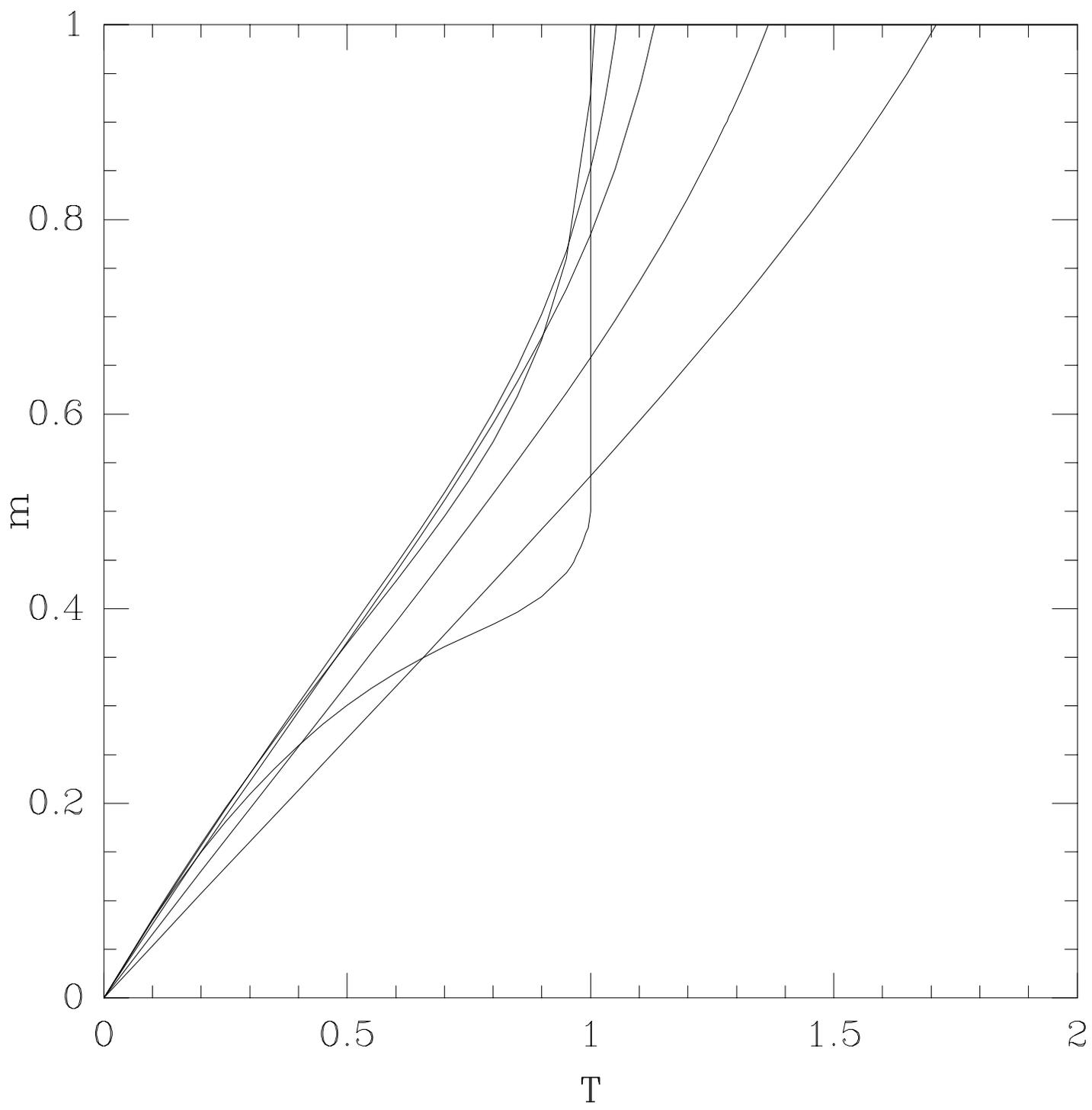
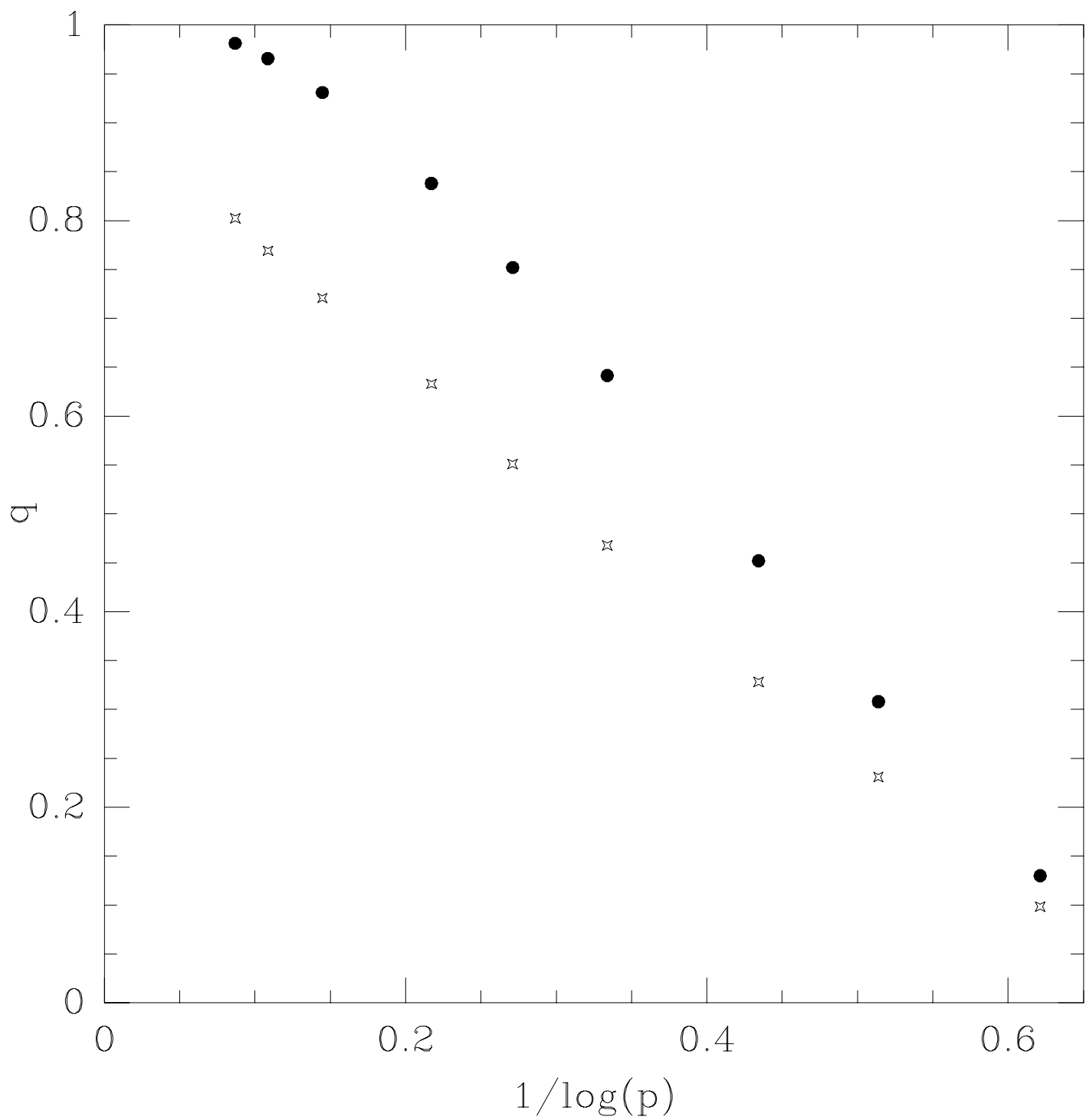
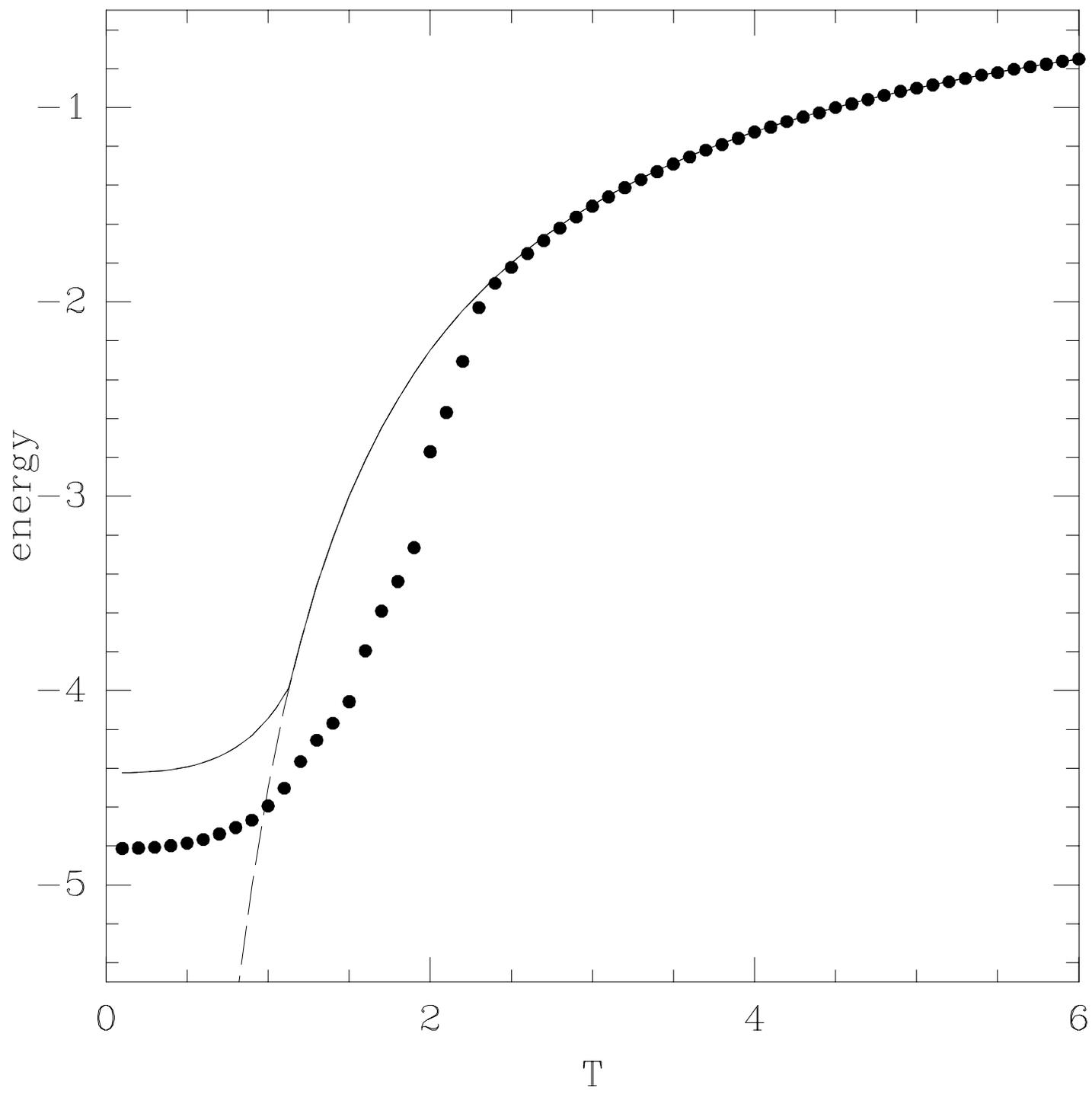
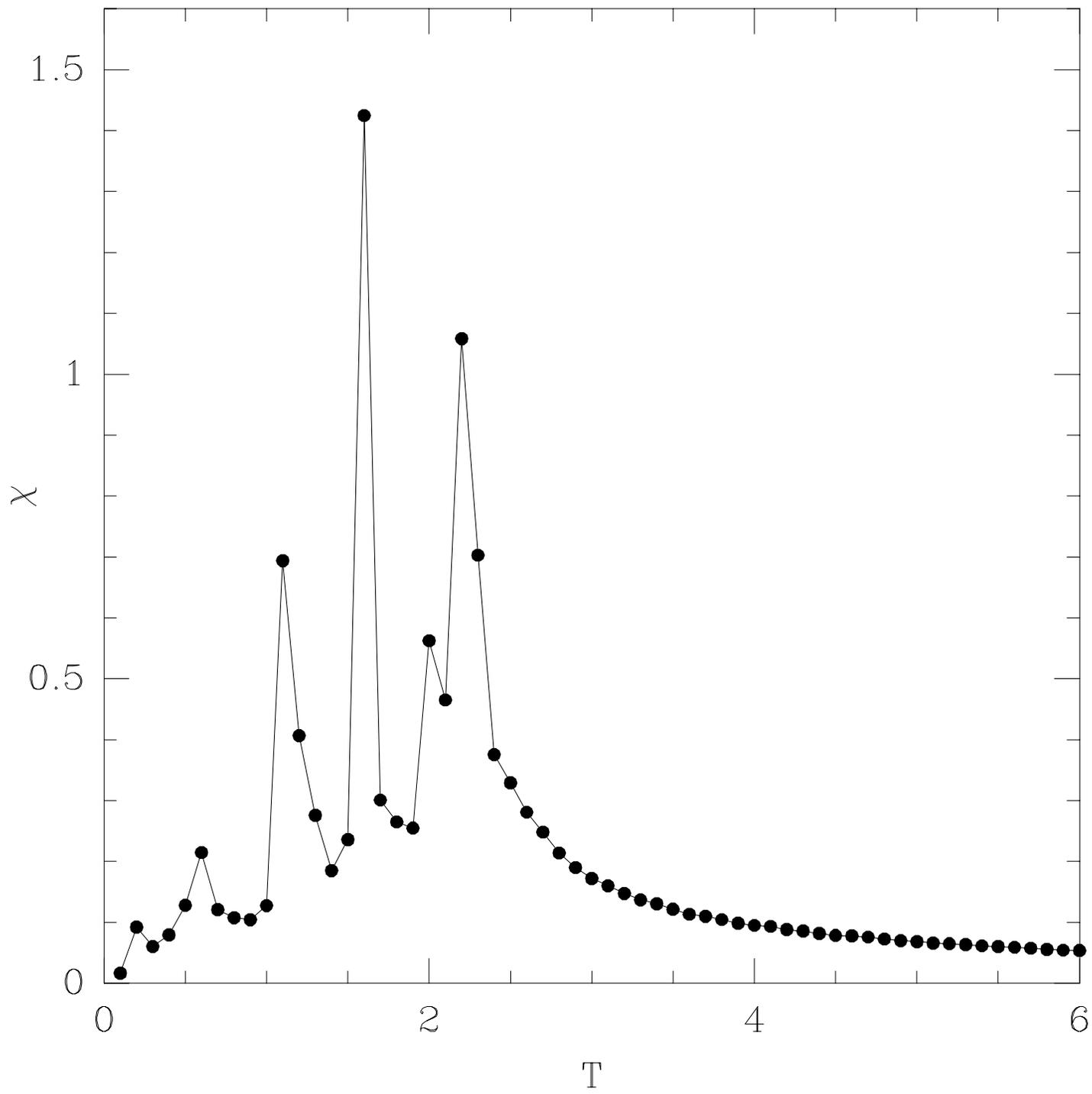
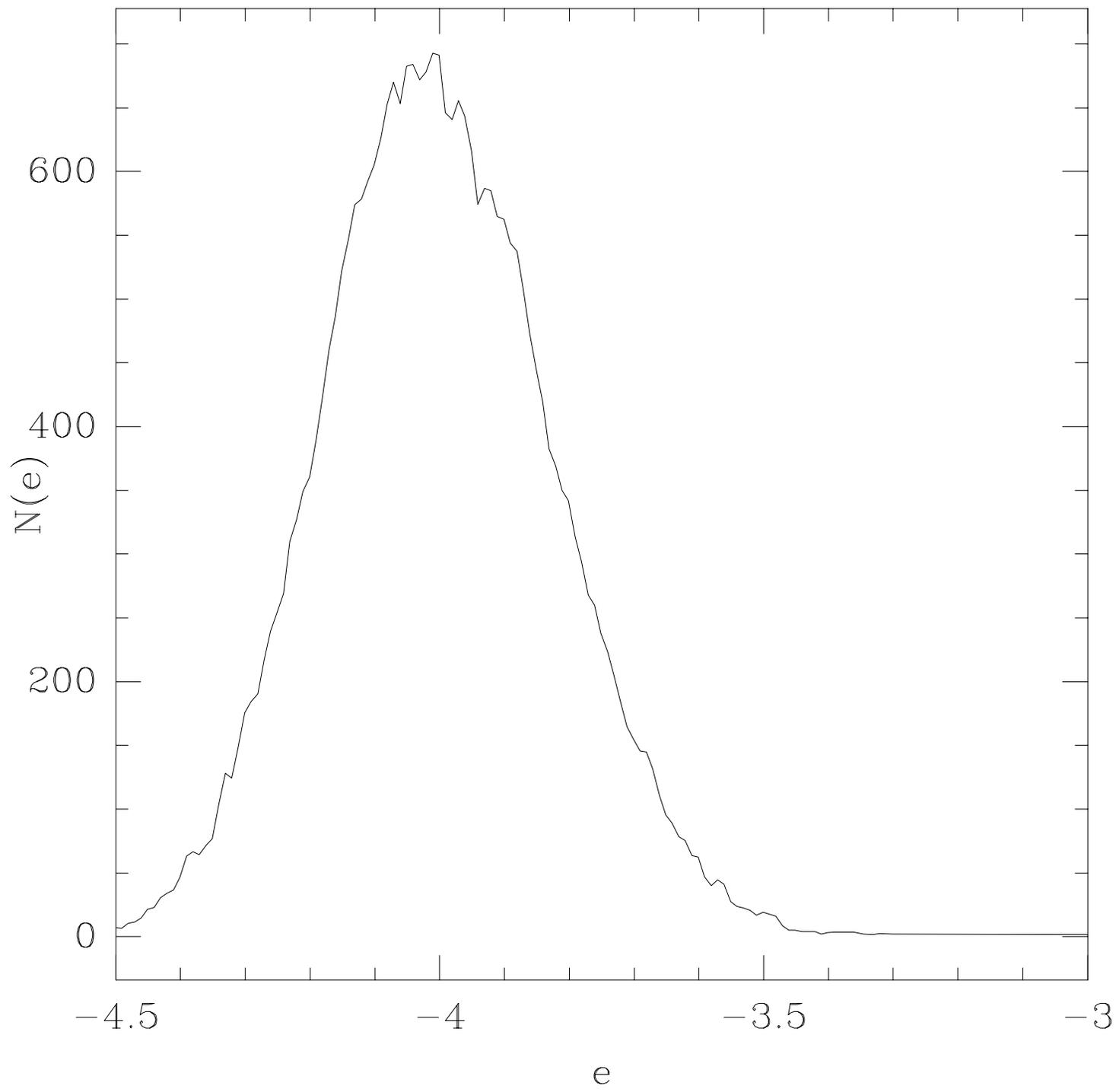
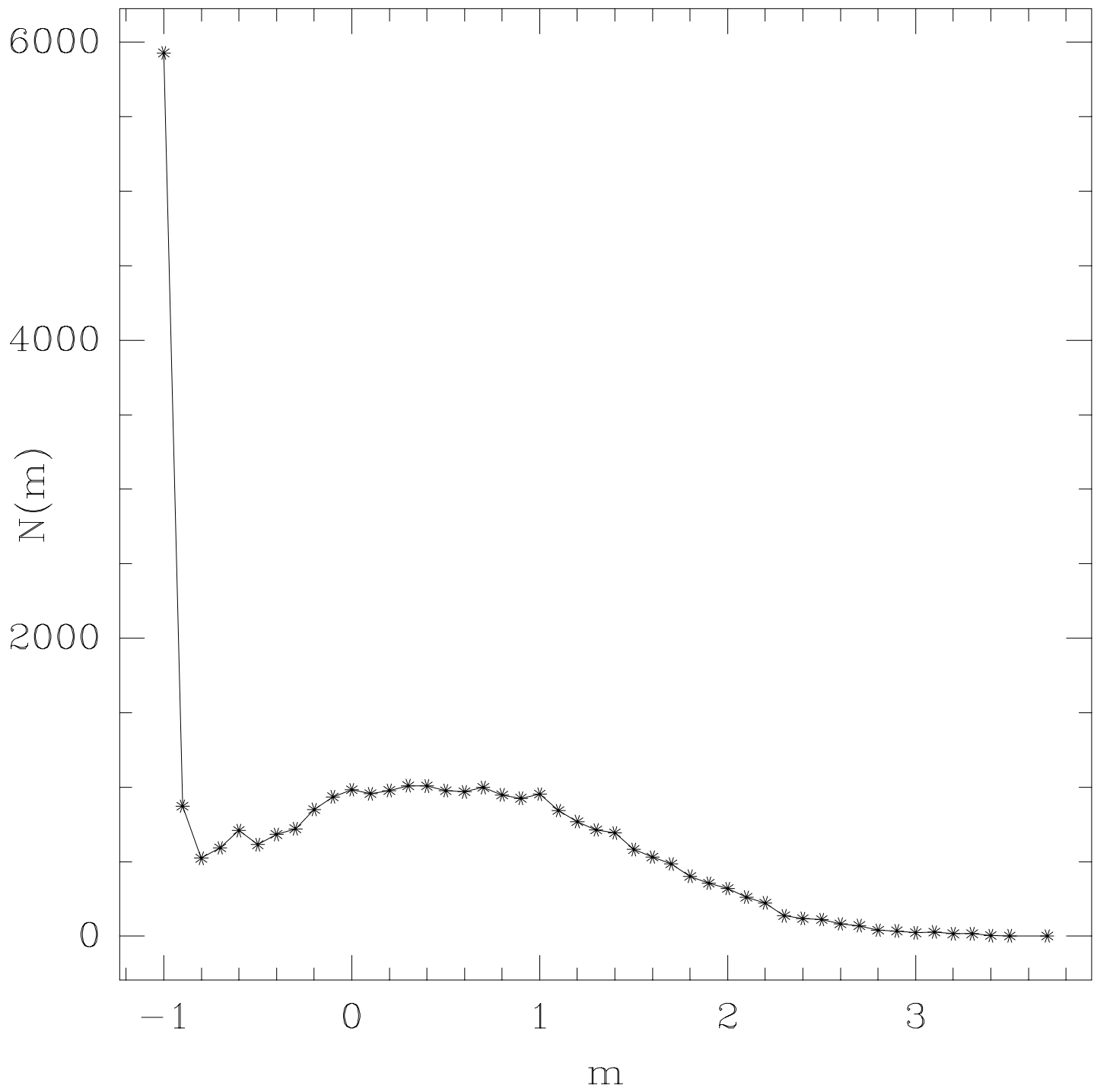
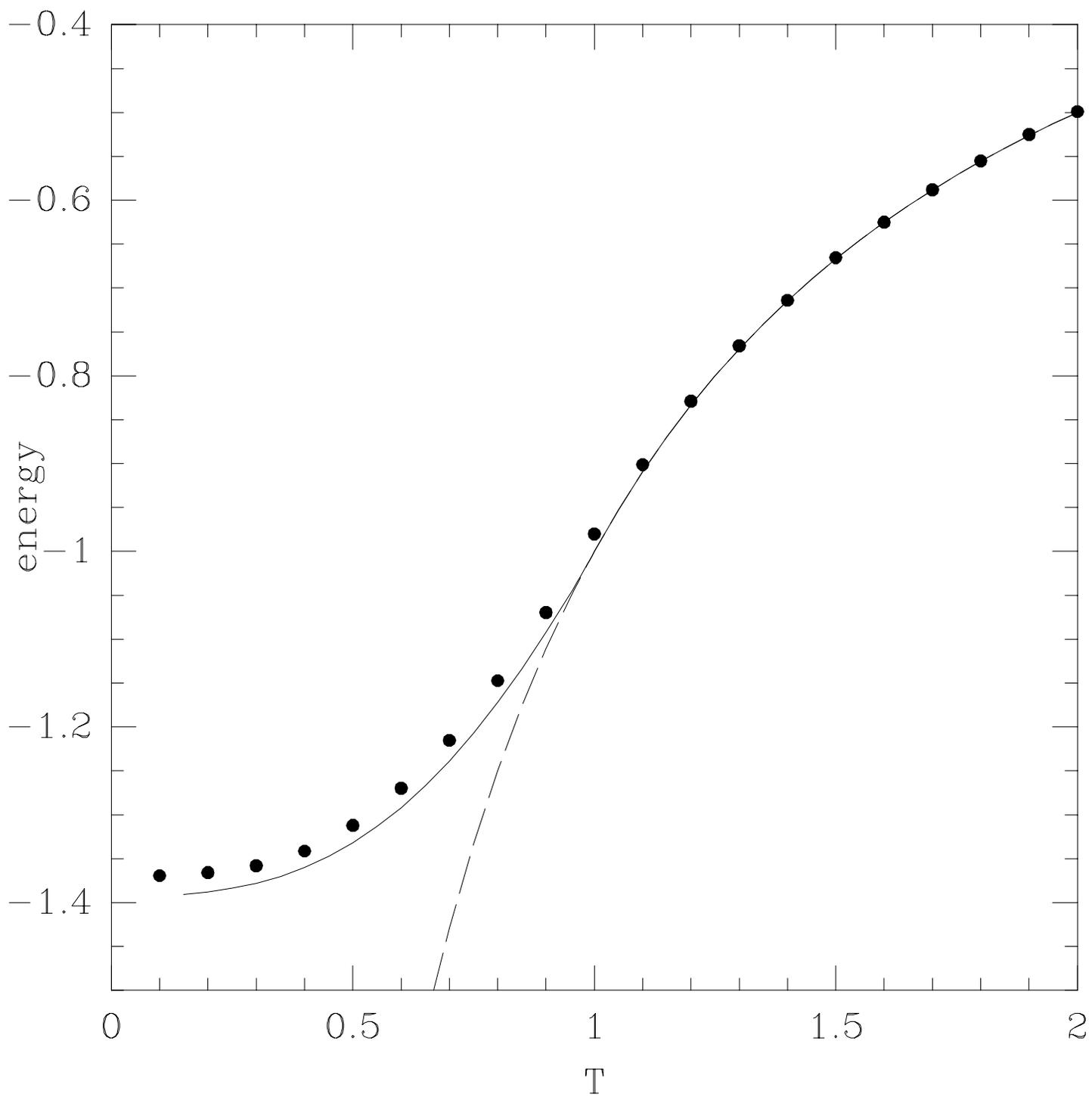
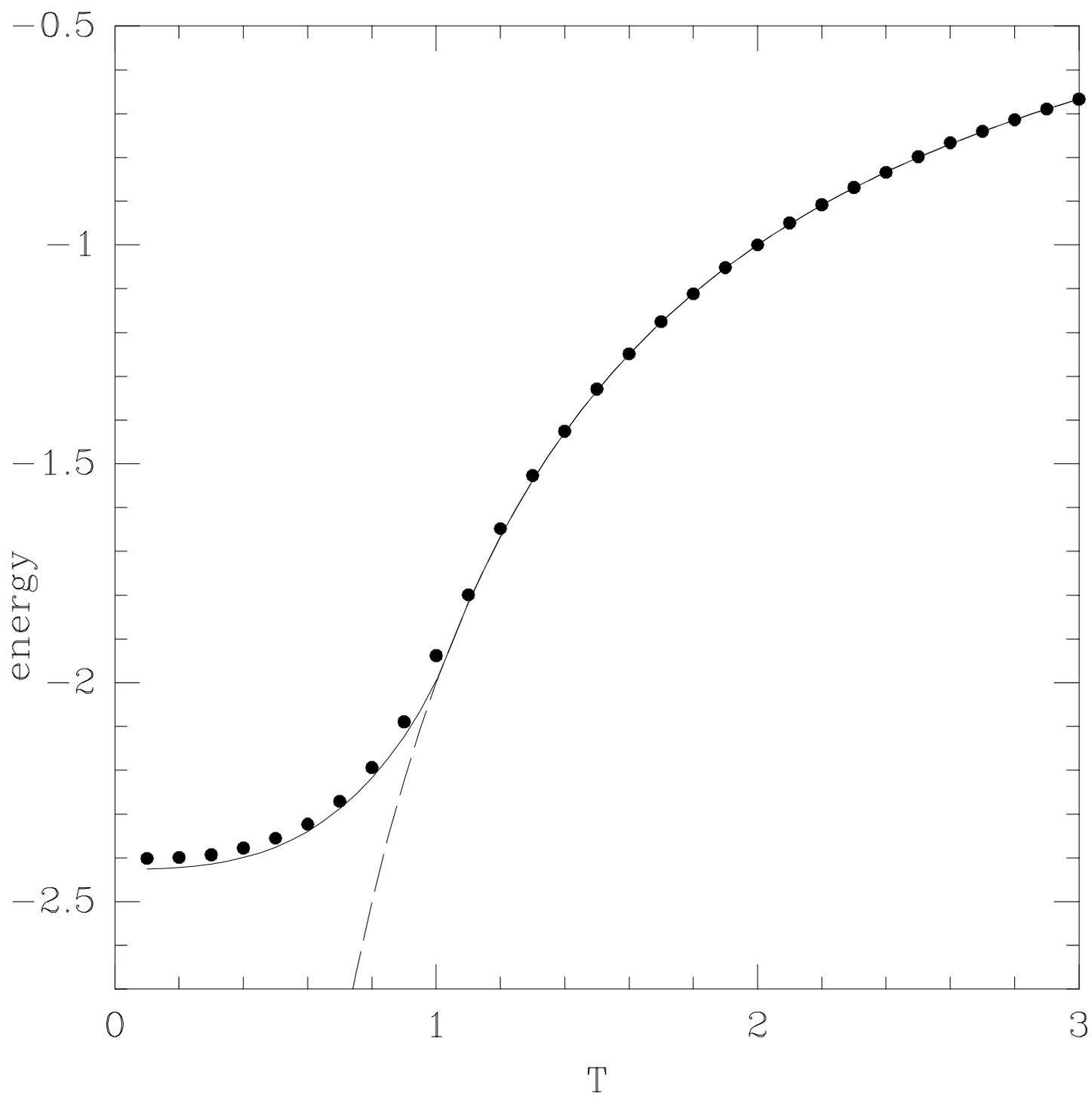
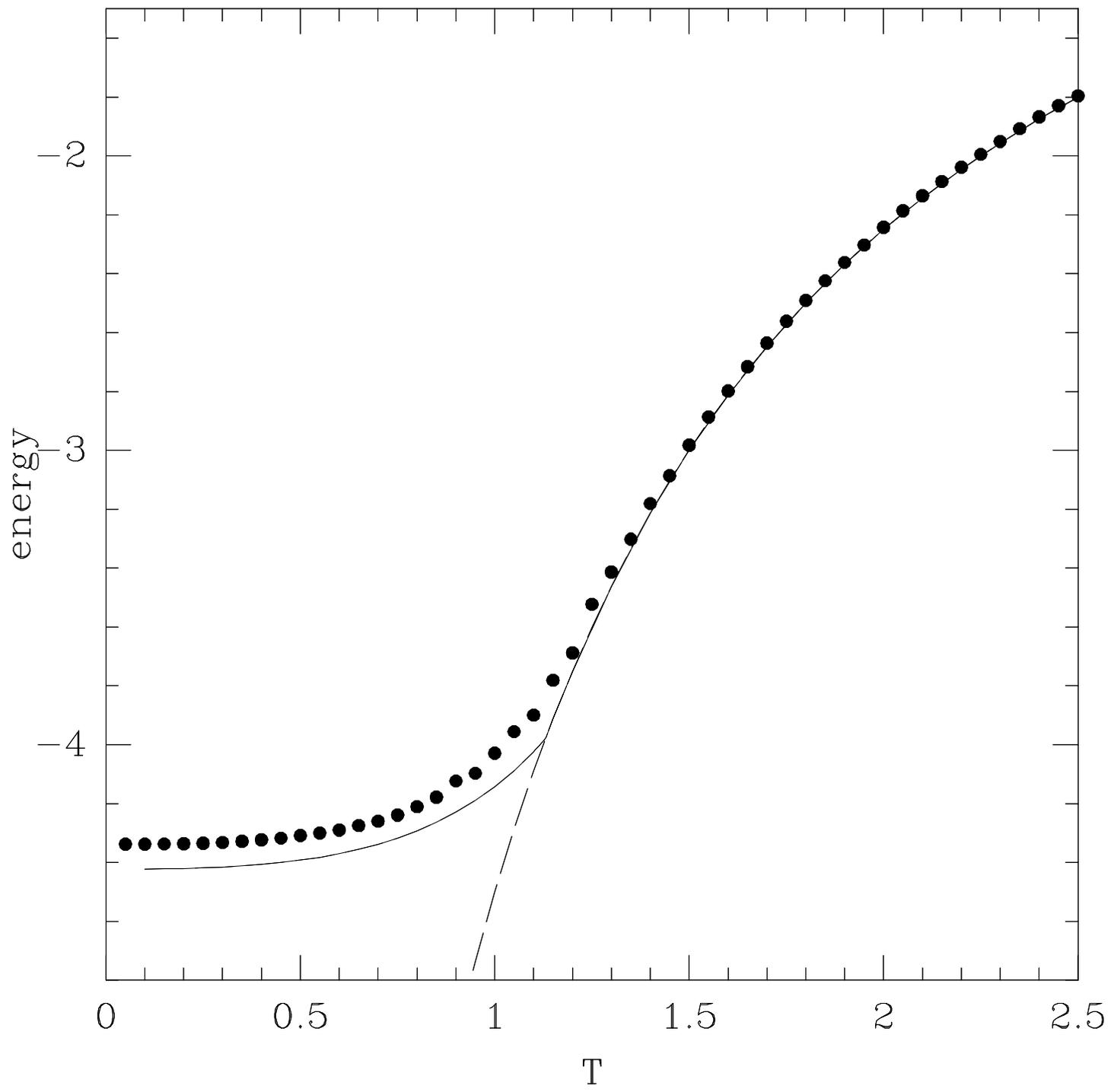
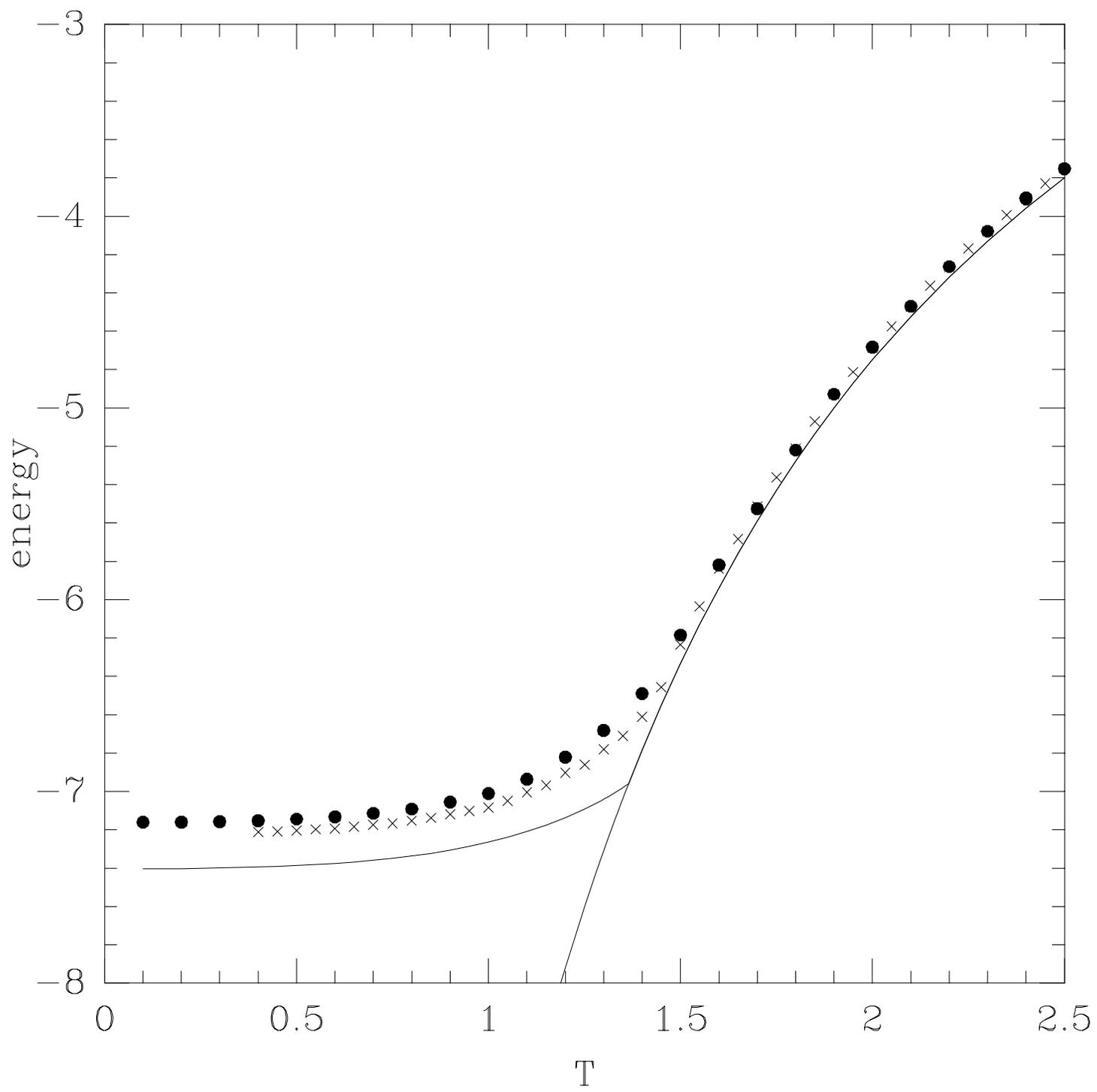
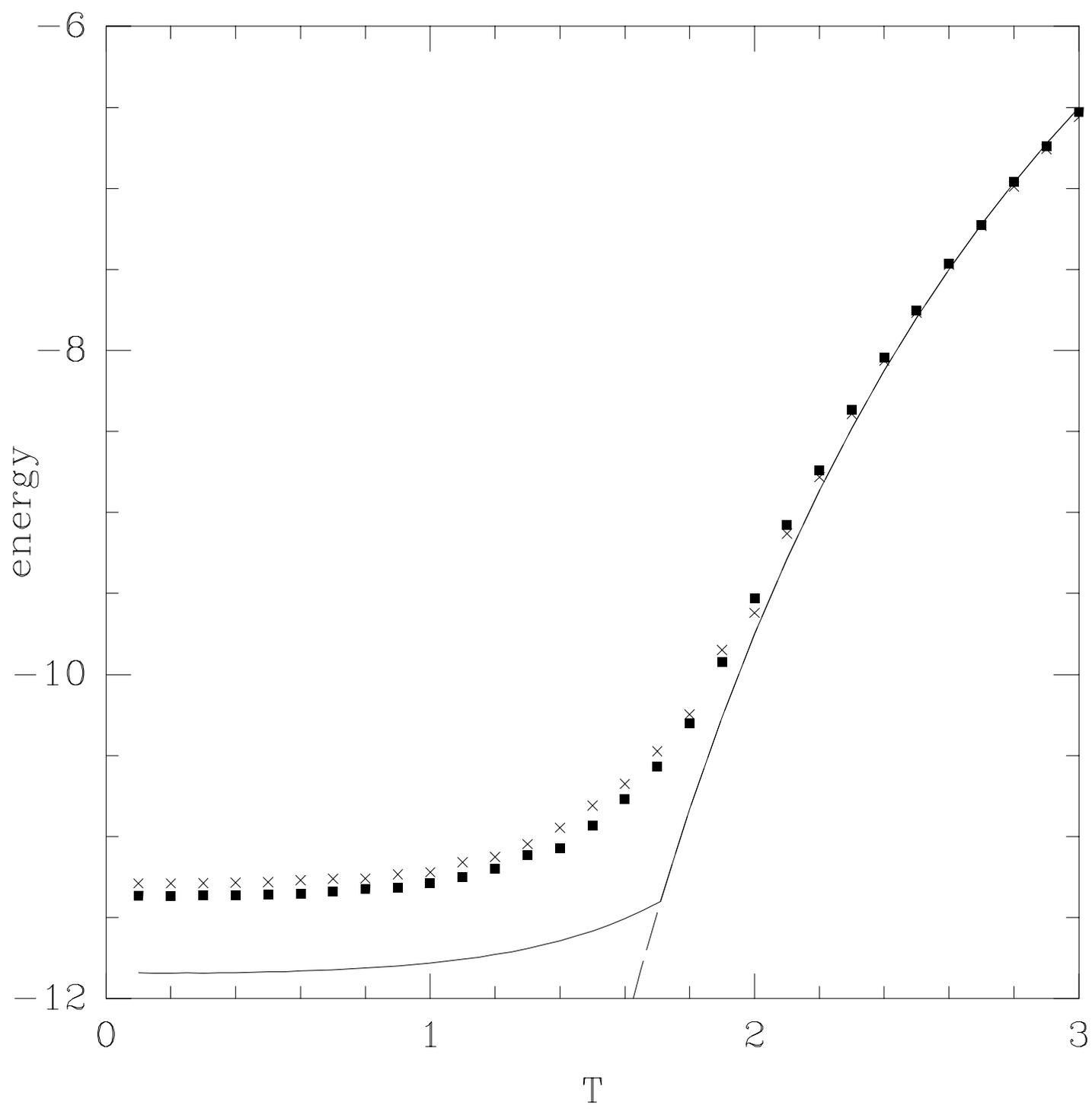
\end{document}